\documentclass[twocolumn,showpacs,preprintnumbers,amsmath,amssymb]{revtex4}
\usepackage{graphicx}% Include figure files
\usepackage{dcolumn}% Align table columns on decimal point
\usepackage{bm}% bold math

\begin{document}

%\preprint{APS/123-QED}

\title{Generalized similarity in finite range solar wind magnetohydrodynamic turbulence}

\author{S. C. Chapman}
\email{S.C.Chapman@warwick.ac.uk}
\author{R. M. Nicol}


\affiliation{Centre for Fusion, Space and Astrophysics, Department of Physics, University of Warwick, Coventry, CV4 7AL, UK}%

\date{\today}% It is always \today, today,

\begin{abstract}
Extended or generalized similarity is a ubiquitous but not well understood feature of turbulence that is realized over a finite range of scales. ULYSSES spacecraft solar polar passes at solar minimum provide \textit{in situ} observations of evolving anisotropic magnetohydrodynamic turbulence in the solar wind under ideal conditions of fast quiet flow. We find a single generalized scaling function characterises this finite range turbulence and is insensitive to plasma conditions. The recent unusually inactive solar minimum -with turbulent fluctuations down by a factor of $\sim 2$ in power- provides a test of this invariance.

\end{abstract}

\maketitle

A characteristic feature of fully developed turbulent flows is an inertial range which is scale invariant in its statistics. Classical signatures of this scale invariance in isotropic homogeneous hydrodynamic turbulence are power law power spectra with a Kolmogorov exponent of $-5/3$ and multiscaling in the structure functions of the velocity field. 
This scale invariance is modified when a characteristic scale becomes apparent, as is the case for measured flows at large but finite Reynolds number which are realized over a finite range of scales. There is longstanding interest in the nature of this correction to ideal hydrodynamic (Kolmogorov) turbulence for example in forced confined flows \cite{moisyprl} and wall bounded shear flow \cite{bchorin}. The nature of this correction captures features of the structure and dynamics of the turbulent flow \cite{sb06,gross94} and when seen from the perspective of finite sized scaling makes contact with critical phenomena \cite{dub}. A ubiquitous, but not well understood, aspect of finite range turbulence is generalised scale invariance or Extended Self Similarity (ESS) \cite{benzi93} which is seen in both hydrodynamic and magnetohydrodynamic (MHD) turbulent flows, for example in the solar wind \cite{carbESS,pagelESS,nicol_08,chapman_09}. In this Letter we find that for solar wind MHD turbulence a single generalized scaling function captures statistical scaling of the largest (outer) scales and is at the heart of the observed ESS, and is insensitive to conditions of the plasma flow. This points to a universal property of finite range MHD turbulence.

The solar wind provides a laboratory to study MHD turbulent fluctuations \cite{brunoR} in a high Reynolds number \cite{matthaeus_05} magnetofluid. \textit{In situ} spacecraft provide single-point plasma observations over a range of scales from seconds to years. Typically, power spectra of solar wind velocity and magnetic field 
components show a power law scaling, suggesting the presence of an inertial range of turbulence and at lower frequencies a ``$\sim 1/f$'' \cite{matthaeus_86,matthaeus_07} spectrum.
 At higher frequencies there is a cross-over to a dissipative or dispersive range which may support a cascade \cite{bale,kiyaniprl09,alex1,alex2,alex3}.  At greater solar distances, and in slow as compared to fast solar wind, this inertial range extends to lower frequencies \cite{horbury_96b} consistent with an actively evolving turbulent cascade. The inertial range turbulence is intermittent with non-Gaussian statistics at small scales \cite{marsch_97,svcastaing,burlaga_02,bruno_07} and anisotropic \cite{milanoprl,chapmangrl,horburyprl} due to a preferred direction introduced by the presence of a background magnetic field, and may also be compressible \cite{hnatprl,carboneprl09}. Ensembles of fluctuations in the ``$1/f$'' region are anisotropic \cite{nicol09} but it is suggested that should conditions of isotropy hold locally, there is evidence for a turbulent cascade on these large scales \cite{svprl07,carboneprl09}. 

Observations in quiet, relatively uniform background magnetic field and flow over extended intervals are provided by the solar polar passes of the ULYSSES spacecraft at times of minimum solar activity. The most recent solar minimum has been reported to be anomalously quiet. Compared to the  previous minimum,  ULYSSES observations show a $\sim-17\%$ lower plasma density and a $\sim 15\%$ lower magnetic field on average; the fluctuations in magnetic field are $\sim 42\%$ lower in power  \cite{mccomas_08,smith_08,issautier_08}. We will compare ULYSSES polar passes for these two sucessive solar minima to test the robustness of generalized scale invariance over different conditions.
We compare $10$ day intervals of $\sim 13,000$ points each of one minute averaged magnetic field component measurements $B_i(t)$ for each of four polar passes, the North and South passes for the two solar minima. The length of these data intervals is sufficient to explore the inertial range scaling and its transition to ``$1/f$'' at lower frequencies.
Contiguous intervals are  selected when ULYSSES is deep within the polar coronal hole flow such that the 
heliospheric distance is  within [$1.92-2.83$ AU] and latitudes  are above $70^{\circ}$ (and peak at $82^{\circ}$) \cite{marsden_96}. The selected observations are of fast quiet polar flows with flow velocity $\sim750$ km/s and are free from large transient coronal events.
We have analyzed in total 24 intervals over the 4 polar passes and find that these all yield the same scaling behaviour in the inertial range that we report here.
 All results are in $RTN$ coordinates: $R$ is along the Sun-spacecraft axis, $T$ is the cross product of $R$ with the Sun's rotation axis and $N$ completes the right-handed system; the ambient field is almost radial. 
Figure 1 compares  a 10 day interval from each of the North polar passes for the sucessive minima at $\sim2.2$ AU. The power spectral density (PSD) for both intervals shows a $\sim -5/3$ exponent power law region with a cross-over to $\sim 1/f$ at lower frequencies, the traces track each other but the overall power level can be seen to differ by a factor of $\sim 2$. 
 We focus on generalized structure functions (GSF) of the field fluctuations $S_p=<\vert B_i(t+\tau)-B_i(t)\vert^{p}>$ where $<...>$ is an ensemble average over $t$ and where in these conditions of fast flow we invoke the Taylor hypothesis and use time interval $\tau$ as our measure of scale. For finite datasets with non-Gaussian Probability Density Functions (PDFs), these statistics can be affected by the presence of large outliers in the PDF tails; we have used the method in \cite{kiyani_06,kiyaniprl07} to verify the robustness of our results against any bias introduced by this source of uncertainty.
Inset in Figure 1 we plot the compensated third order structure functions $S_3/\tau$. 
These curves have different amplitudes due to different power levels in the fluctuations but intriguingly
  all roughly have the same functional form.

\begin{figure}
\includegraphics[scale=0.3]{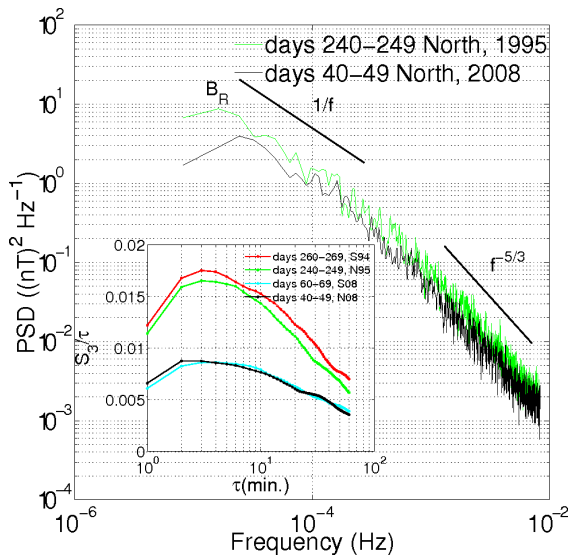}
\caption{Log-log plot of the \textit{R} magnetic field component PSD for $10$ day intervals from ULYSSES North polar passes in $1995$ and $2008$. ULYSSES was at a similar heliospheric distance for these intervals. There is a difference in power of $\sim-42\%$ between the two minima. Inset: Compensated GSF $S_{3}/\tau$ (linear scale) versus $\tau$ (logarithmic scale), for the \textit{R} component for all passes at this heliospheric distance.}
\end{figure}

Let us now examine the GSF in more detail. In the main panel of Figure 2 we plot $S_3$ versus $\tau$ for six consecutive 10 day intervals. 
The inertial range extends from below the smallest ($1$ minute) timescale up to $\tau \sim 20 $ minutes (see also \cite{horbury_96b,nicol_08}), and we can see that over this range there is no single straight line fit, as would be the case for fully developed turbulence for which one anticipates the scaling $S_p \sim \tau^{\zeta(p)}$. Extended Self Similarity \cite{benzi93} instead suggests the scaling of ratios of the GSF:
\begin{equation}
S_{p}(\tau) \sim [S_q(\tau)]^{\zeta(p)/\zeta(q)}
\end{equation}
and we test this in the inset of Figure 2 where we plot $S_3$ versus $S_2$. We can see that ESS indeed captures the scaling of the inertial range (but does not extend into the  ``$1/f$'' range, for a detailed ESS analysis of the 1995 polar pass see \cite{nicol_08}). We find that ESS holds for the inertial range of all of the intervals under study.
\begin{figure}
\includegraphics[scale=0.3]{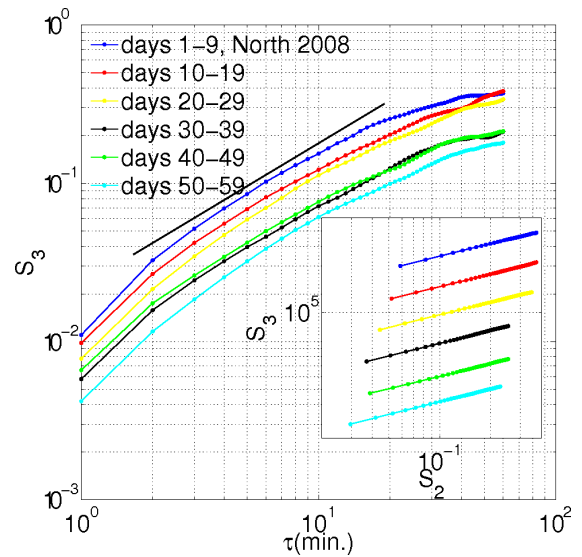}
\caption{GSFs of $S_{3}$ versus $\tau$ on log-log axes for the $R$ component of the magnetic field for $5$ consecutive $10$ day intervals from the North polar pass of $2008$ shown for $\tau=2-60$ minutes. The inset figure shows ESS plots of $S_{3}$ against $S_{2}$ on logarithmic axes, the different intervals are shifted for clarity.}
\end{figure}
ESS implies a generalised similarity of the form:
\begin{equation}
S_p(\tau)=S_p(\tau_0)[g(\tau/\tau_0)]^{\zeta(p)} 
\end{equation}
however the observation of ESS alone does not discern whether or not the function $g(\tau/\tau_0)$ is ubiquitous in nature or whether it varies across different components of the magnetic field or with ambient conditions. We first establish that the same function $g$ is obtained for different orders $p$ (as required by ESS) and also for all three components of the field and for each of the polar passes. To obtain an expression for $g(\tau/\tau_0)$ we need to invert equation (2) however since we do not have the scaling $S_p \sim \tau^{\zeta(p)}$ we cannot determine the exponents $\zeta(p)$ directly. Instead we rearrange equation (2) to give:
\begin{equation}
\bar{g}(\tau,\tau_0)=\left[ \frac{S_p(\tau)}{S_p(\tau_0)}\right]^{\frac{\zeta(3)}{\zeta(p)}}=g(\tau/\tau_0)^{\zeta(3)} 
\end{equation}
ESS plots of one GSF against another (as in the inset of Figure 2) then yield measurements of the ratios of the exponents $\zeta(3)/\zeta(p)$. 
We plot $\bar{g}$ versus $\tau$ on linear axes in Figure 3 for orders $p=1-4$ and for all of the magnetic field components. Each of the four panels refers to a 10 day interval from each of the polar passes, again at the same heliospheric distance.
The plotted range of 
$\tau=1-60$ minutes encompasses the inertial range and the cross-over to the ``$1/f$'' range. The normalization scale $\tau_0$ is arbitrary and we used
 $\tau_{0}=10$ minutes which is within the inertial range for all the intervals. In all four panels of Figure 3 we can see two distinct regions: an inertial range at smaller $\tau$ where the curves closely correspond to each other, and the ``$1/f$''  range at larger $\tau$, where the curves diverge. Variability in the scaling exponent $\alpha$ in the '$1/f^\alpha$" , $\alpha \sim 1$ region has been noted previously, both with ambient conditions \cite{horbury_95,horbury_96,nicol09} and with field component \cite{nicol_08}.
Figure 3 demonstrates that there is a robust (i.e. usually present) inertial range which follows a generalized scaling captured by function $\bar{g}(\tau)$ which satisfies ESS (i.e. it is the same for all $p$). Intriguingly, it is also the same 
 for all of the field components, despite the differences in their power levels, and we have verified that this is the case for all of the intervals under study.
 The outer scale $\tau_c$ where the inertial range terminates
 varies between all four polar passes and is not simply ordered by the distinct ambient conditions of the two solar minima; the longest ($\tau_c \sim 25$ minutes) and the shortest ($\tau_c \sim 12 $ minutes) both occurring respectively for the 1994 and 2007 South solar minima.

\begin{figure}
\includegraphics[scale=0.23]{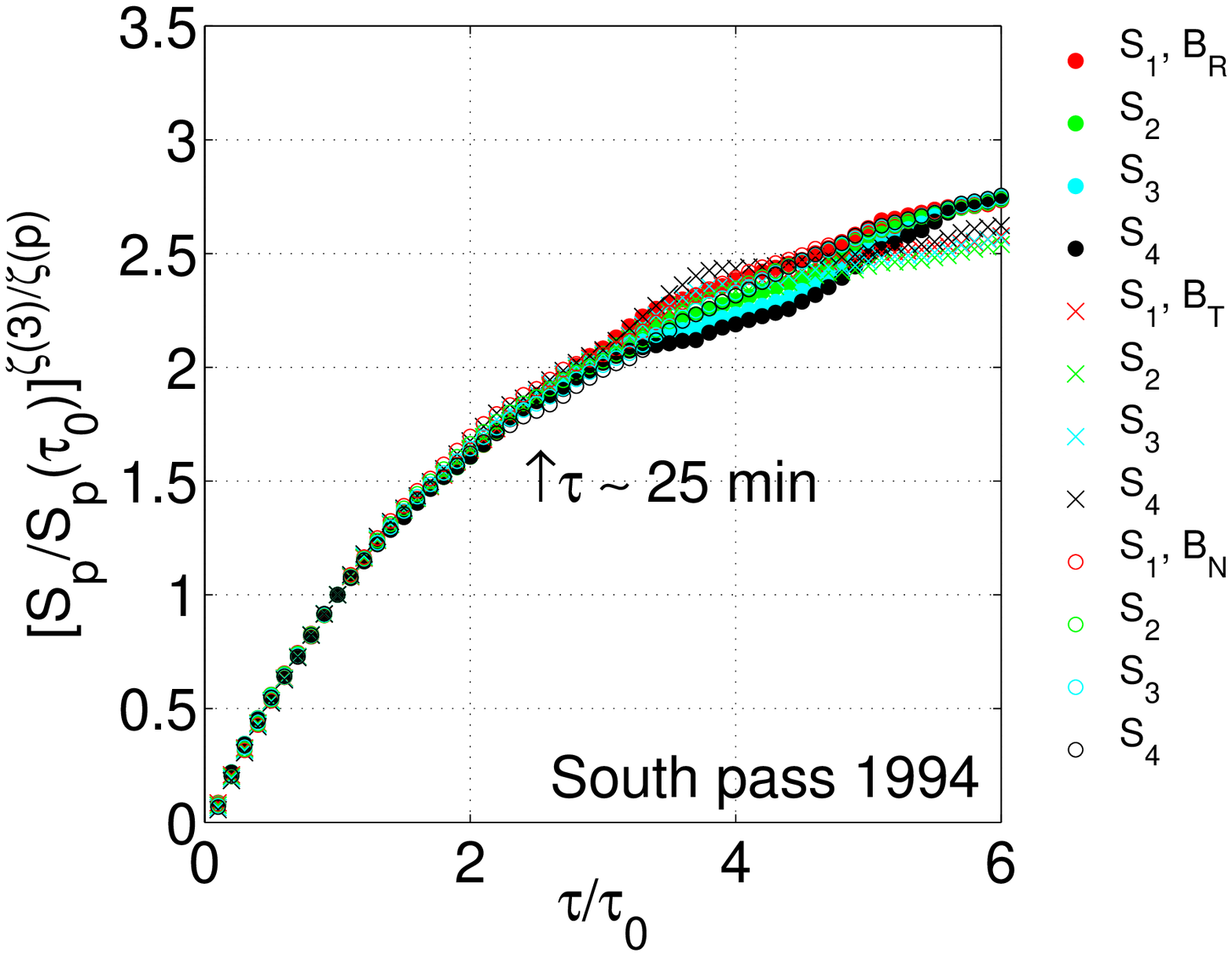}
\includegraphics[scale=0.23]{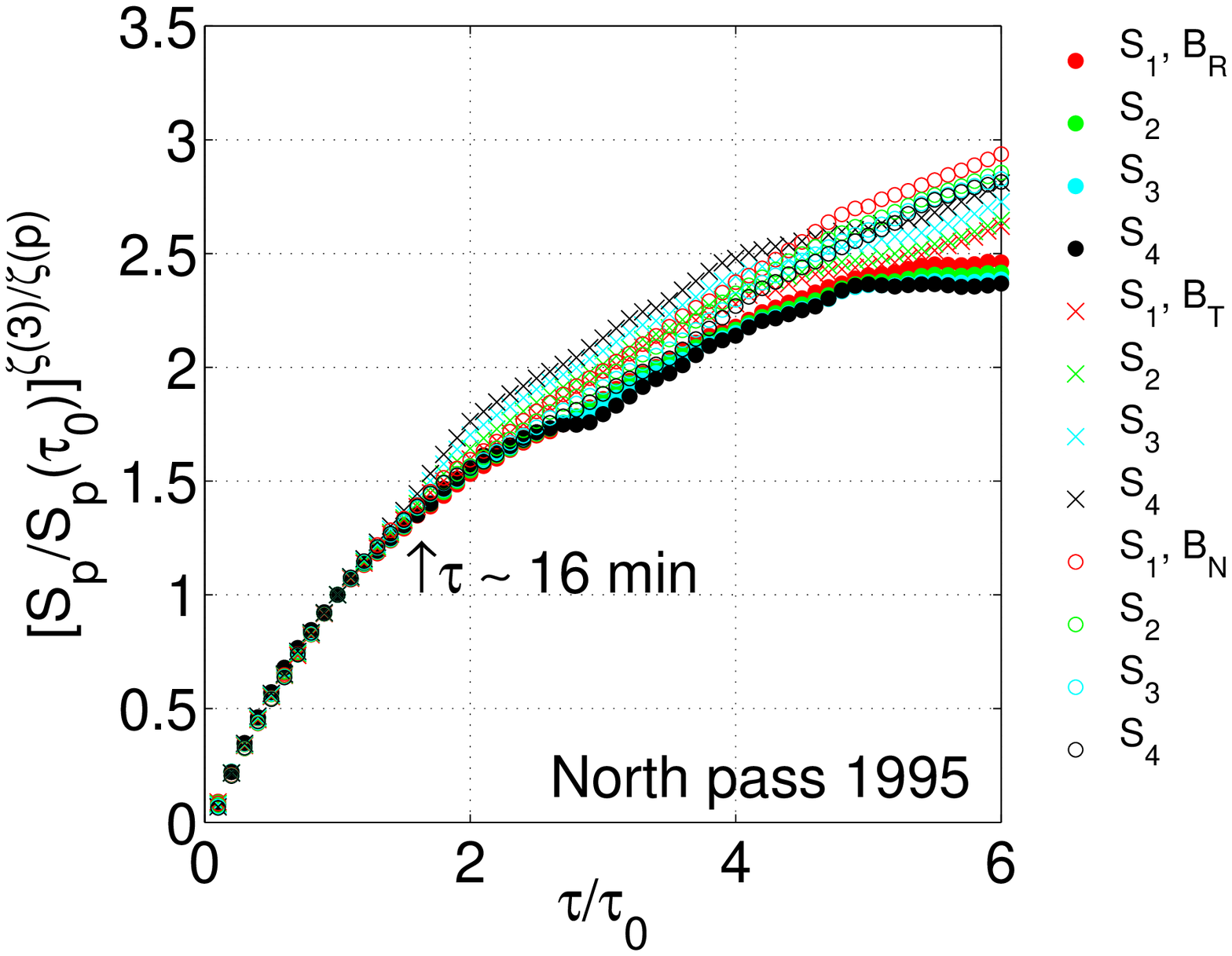}
\includegraphics[scale=0.23]{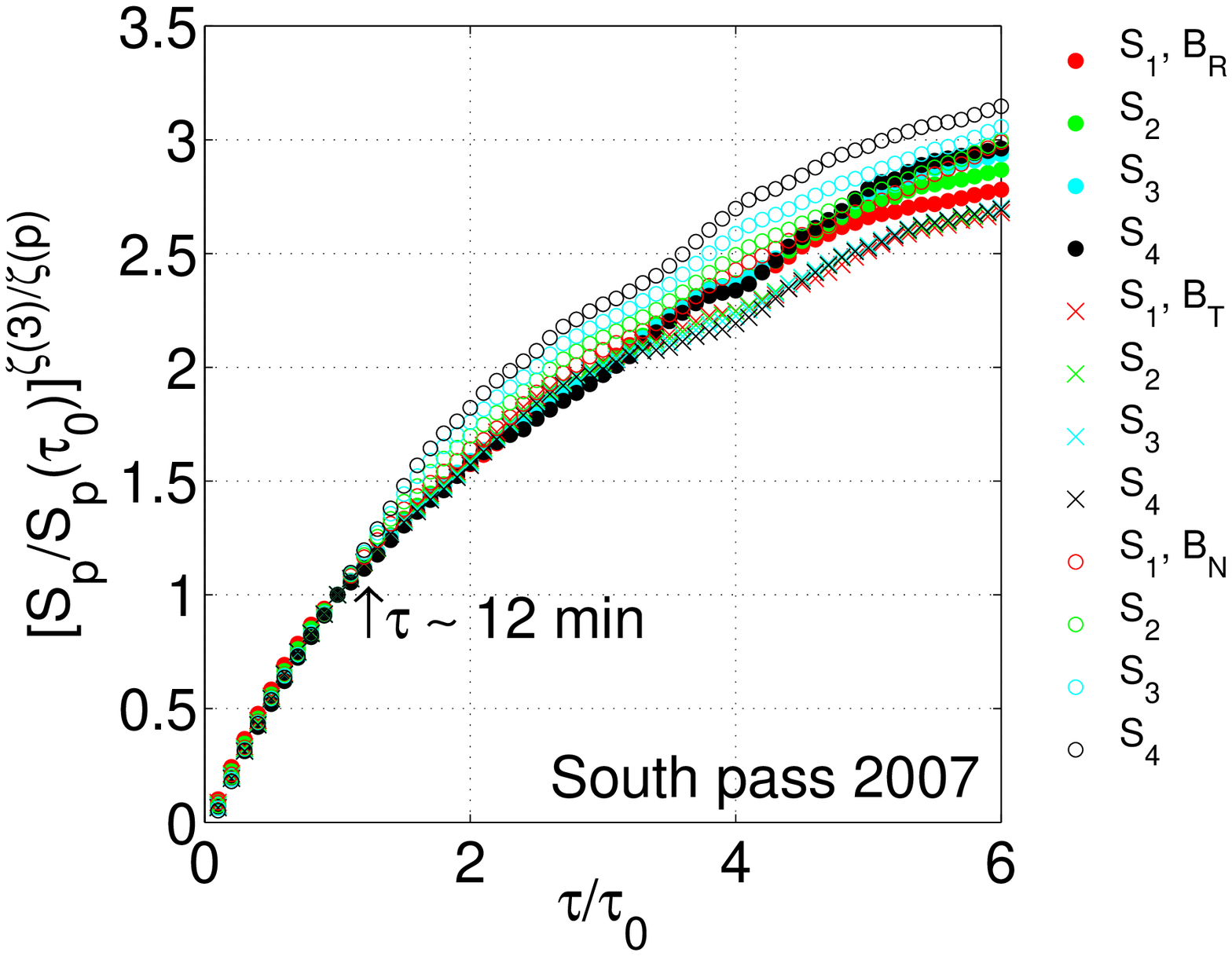}
\includegraphics[scale=0.23]{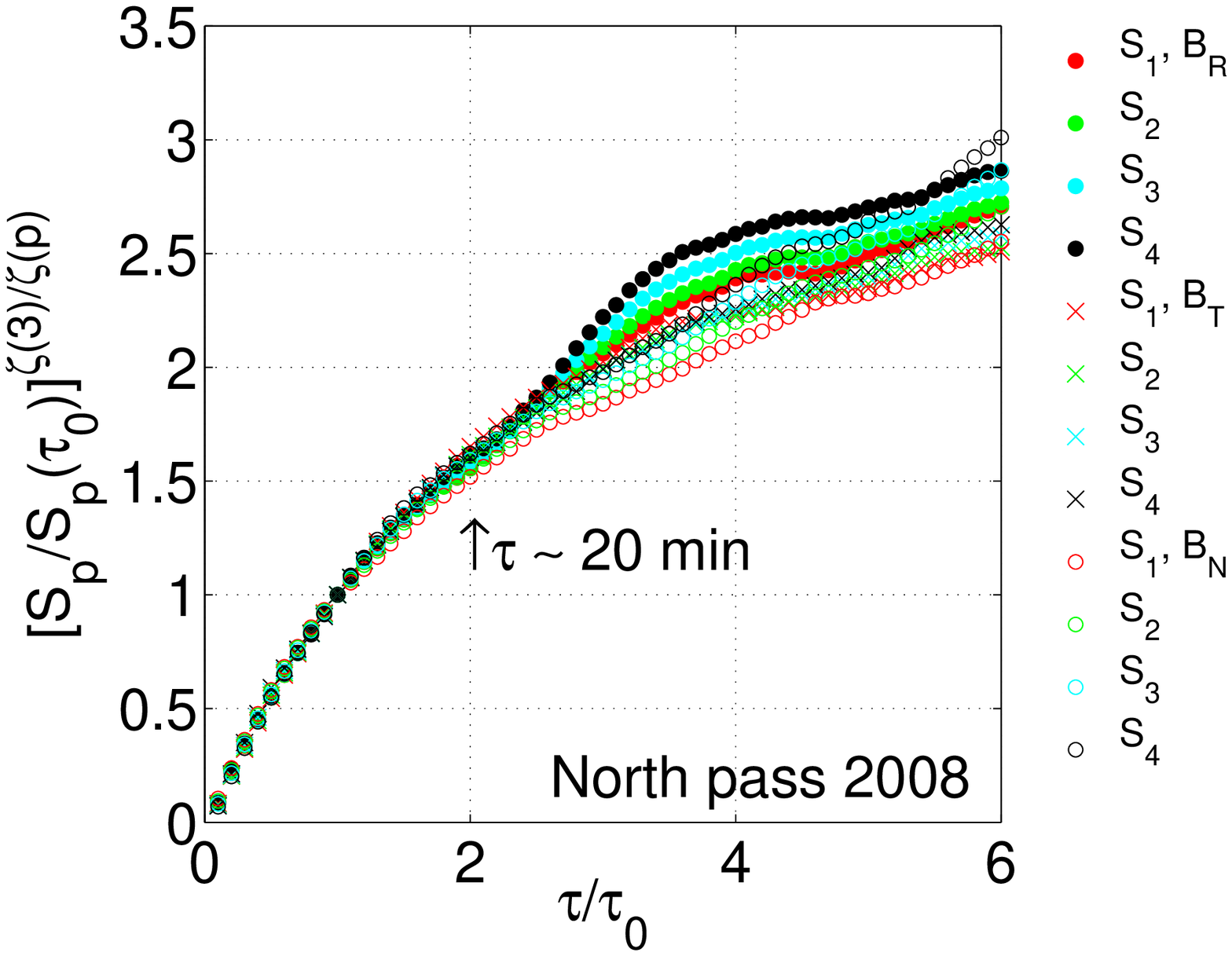}
\caption{$[S_p(\tau)/S_p(\tau_0)]^{\zeta(3)/\zeta(p)}$ versus $\tau/\tau_0$, ($\tau_0=10$ minutes) for $p=1$ to $4$ for all field components are overplotted for $10$ day intervals at the same heliocentric distance for each of the ULYSSES polar passes: South $1994$ (days $260-269$), North $1995$ (days $240-249$), South $2007$ (days $60-69$) and North $2008$ (days $40-49$).}
\end{figure}
We next test whether the same function $\bar{g}(\tau)$ holds for all of the polar passes. Figure 4 again plots 
$\bar{g}(\tau)$ versus $\tau$ on linear axes, now for $p=3$ for all field components for all four polar passes shown in Figure 3. Again we find an inertial range where there is a close correspondence between all these curves. A good fit to this function is of the form $h(\tau,\tau_0,\hat{\tau})=a.(\tau/\hat{\tau})^{({\tau/\tau_{0})}^b}$ where we have chosen 
scale parameters $\tau_{0}=10$ minutes  and $\hat{\tau}=1$ minute and obtained constant fitted parameters $a=0.101\pm0.001$ and $b=0.10\pm0.01$, this is overlaid on the figure.
This single function $\bar{g}$ reflects both the underlying phenomenology of the inertial range MHD turbulence in the value of the exponent $\zeta(3)$ and the (generalized) scale invariance of the largest structures in the evolving turbulence in the function $g$. To recover the \textit{same } $\bar{g}(\tau)$ over this variety of ambient conditions suggests invariant phenomenology for \textit{both} these aspects of this finite range, evolving MHD turbulence. Although we cannot precisely determine $\zeta(3)$ the robustness of the inertial range would suggest that it does not change significantly between the different intervals. If so, we have shown an important property of MHD turbulence as seen evolving in the solar wind- that ESS arises as a consequence of a single generalized similarity of finite range turbulence that is invariant to ambient conditions and the level of power in the turbulent signal. 

\begin{figure}
\includegraphics[scale=0.45]{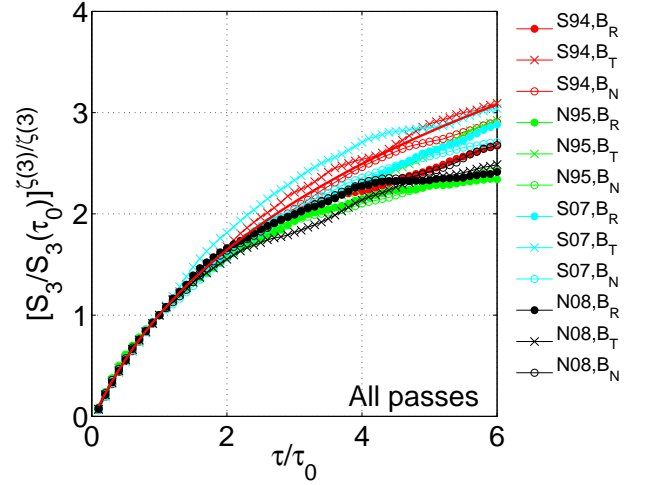}
\caption{$[S_3(\tau)/S_3(\tau_0)]^{\zeta(3)/\zeta(3)}$ versus $\tau/\tau_0$ for all field components for the $10$ day intervals plotted in Figure 3.}
\end{figure}

To gain some insight into the robustness of this generalized similarity we finally turn to the  PDFs of the fluctuations $B_i(t+\tau)-B_i(t)$. Figure 5 shows the
PDFs for the radial field component fluctuations in the inertial range ($\tau=6$ minutes shown) for all of the intervals in Figure 3, normalized to $\sigma$ to afford a comparison of the functional form. We can see that all of the curves collapse onto each other - the functional form of the turbulent fluctuations is invariant to different power levels in the turbulence and different ambient conditions. The PDF is non- Gaussian (a Gaussian is fitted for comparison), a well-known feature of small scale solar wind turbulence, and is stretched exponential in form. All three components show this invariance to ambient conditions in the functional forms of their PDFs. 
\begin{figure}
\includegraphics[scale=0.40]{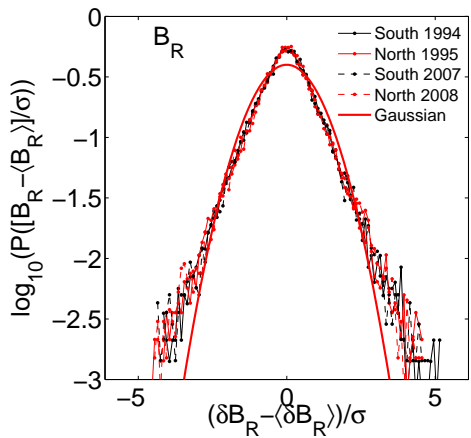}
\caption{The PDF of fluctuations in the \textit{R} magnetic field component on timescale $\tau=6$ minutes for $10$ day intervals plotted in Figure 3. A Gaussian fit to the data is also plotted. }
\end{figure}
The field components differ in the power levels of their fluctuations and this is captured by the standard deviation $\sigma$. For the $R$ component $\sigma = [0.37-0.3] nT \pm 0.002$ whereas for the $T$ and $N$ components $\sigma = [0.52-0.42] nT \pm 0.003$ between the 1994/5 and 2007/8 solar minimum passes. For comparison, the mean magnetic field is $\sim1.4$ nT (94/95 polar pass) and $\sim1$ nT (07/08 polar pass). There is therefore a consistent anisotropy in the amplitude of fluctuations between the $R$ and the $T,N$ components, present for both solar minima as well as an overall decrease in power between minima.

The recent unusually quiet solar minimum provides a unique opportunity to make a relatively controlled comparison of the statistical scaling 
 of evolving MHD turbulence \textit{in situ} in the quiet, fast solar wind under different ambient plasma parameters (density, magnetic field magnitude) and in the power level of the turbulence. We find evidence of the same generalized scaling in all 24 intervals that we  consider. The robustness of this generalized similarity  suggests that this may be a universal property of finite range anisotropic MHD turbulence. How this relates to the statistical scaling properties of the evolving solar wind under all manifestations- such as in slow as opposed to fast flow, and  over the solar cycle \cite{kiyaniprl07}- is an open question.
 Opportunities to test for this generalized similarity in finite range turbulence are also in the turbulent foreshock \cite{naritaprl} and in the magnetosheath \cite{sundprl} provided sufficiently long stationary intervals can be identified. 
\begin{acknowledgments}
We acknowledge the EPSRC, the STFC and CCFE for financial support and A. Balogh and the ULYSSES team for data provision. We thank E. Leonardis and K. Kiyani for discussions.
\end{acknowledgments}


\begin{thebibliography}{20}
\bibitem{moisyprl} F. Moisy, P. Tabeling,  H. Willaime, Phys. Rev. Lett, 82, 3994, (1999)
\bibitem{bchorin} G. I Barenblatt, A. J. Chotin, PNAS, 101, (2004)
\bibitem{sb06} K. R. Sreenivasan, A. Bershadskii, J. Fluid Mech., 554, 477, (2006)
\bibitem{gross94}S. Grossmann, D. Lohse, V. L'vov, I. Procaccia, Phys. Rev. Lett, 73, 432, (1994)
\bibitem{dub} B. Dubrulle, Eur. Phys. J. B 14, 757 (2000)
\bibitem{benzi93} R. Benzi et al, Phys. Rev. E 48, R29 (1993).
\bibitem{carbESS}V. Carbone, R. Bruno, P. Veltri, Geophys. Res. Lett., 23, 121, (1996)
\bibitem{pagelESS}C. Pagel, A. Balogh, Nonlin. Proc. Geophys. 8, 313, (2001) 
\bibitem{chapman_09} S. C. Chapman et al, ApJ Lett. 695, L185 (2009).
\bibitem{nicol_08} R. M. Nicol, S. C. Chapman, R. O. Dendy, Ap. J., 679,862 (2008)
\bibitem{brunoR} R. Bruno, V. Carbone, Living Rev. Solar Phys. 2, 4 (2005)
\bibitem{matthaeus_05} W. H. Matthaeus et al, Phys. Rev. Lett., 95, 231101 (2005).
\bibitem{matthaeus_86} W. H. Matthaeus, M. L. Goldstein, Phys. Rev. Lett., 57, 495 (1986).
\bibitem{matthaeus_07} W. H. Matthaeus et al, ApJ Lett. 657, L121, (2007).
\bibitem{kiyaniprl09}K. H. Kiyani et al, Phys. Rev. Lett., 103, 075006 (2009)
\bibitem{alex1}O. Alexandrova , V.  Carbone, P. 
Veltri, L. Sorriso-Valvo, Planet. Space Sci., 55, 2224, (2007).
\bibitem{alex2}O. Alexandrova, V.  Carbone, P. 
Veltri, L. Sorriso-Valvo, Ap. J,   674, 1153, (2008)
\bibitem{alex3}O. Alexandrova et al, Phys. Rev. Lett., 103, 165003, (2009)
\bibitem{bale} S. D. Bale et al, Phys. Rev. Lett. 94, 215002 (2005).

\bibitem{horbury_96b} T. S. Horbury, A. Balogh, R. J. Forsyth, E. J. Smith, Astron. Astrophys. 316, 333 (1996).

\bibitem{bruno_07} R. Bruno et al, Phys. Plasmas, 14, 032901 (2007).
\bibitem{marsch_97} E. Marsch, C.-Y. Tu, Nonlin. Proc. Geophys., 4, 101 (1997).
\bibitem{svcastaing} L. Sorriso-Valvo et al, Planet. Space Sci. 49, 1193 (2001).
\bibitem{burlaga_02} L. F. Burlaga, M. A. Forman, J. Geophys. Res. 107, 1403 (2002).
\bibitem{milanoprl}L. J. Milano, S. Dasso, W. H. Matthaeus, C. W. Smith, Phys. Rev. Lett, 93, 155005, (2004)
\bibitem{chapmangrl} S. C. Chapman, B. Hnat, Geophys. Res. Lett. 34, 17103 (2007).
\bibitem{horburyprl} T. S. Horbury, M. Forman, S. Oughton, Phys. Rev. Lett., 101, 175005 (2008)

\bibitem{hnatprl} B. Hnat, S. C. Chapman,  G. Rowlands, Phys. Rev. Lett. 94, 204502 (2005).
\bibitem{carboneprl09} V. Carbone, et al, Phys. Rev. Lett. 103, 061102 (2009)
\bibitem{nicol09} R. M. Nicol, S. C. Chapman, R. O. Dendy, Ap. J., 703, 2138, (2009)

\bibitem{svprl07} L. Sorriso-Valvo, et al, Phys. Rev. Lett., 99, 115001 (2007) 
\bibitem{mccomas_08} D. J. McComas, et al., Geophys. Res. Lett. 35, 18103 (2008).
\bibitem{smith_08} E. J. Smith, A. Balogh, Geophys. Res. Lett. 35, 22103 (2008).
\bibitem{issautier_08} K. Issautier et al, Geophys. Res. Lett. 35, 19101 (2008).
\bibitem{marsden_96}R. G. Marsden, E. J. Smith, J. F. Cooper,  C. Tranquille, Astron. Astrophys. 316, 279 (1996).
 

\bibitem{kiyani_06} K. Kiyani, S. C. Chapman, B. Hnat, Phys. Rev. E 74, 051122 (2006)
\bibitem{kiyaniprl07}  K. Kiyani, S. C. Chapman, B. Hnat, R. M. Nicol, Phys. Rev. Lett. 98, 211101 (2007)
\bibitem{horbury_95} T. S. Horbury, A. Balogh, R. J. Forsyth, E. J. Smith, Geophys. Res. Lett. 22, 3401 (1995).
\bibitem{horbury_96} T. S. Horbury, A. Balogh, R. J. Forsyth, E. J. Smith, Astron. Astrophys. 316, 333 (1996).
\bibitem{naritaprl}Y. Narita, K.-H. Glassmeier, R. A. Treumann, Phys. Rev. Lett. 97, 191101 (2006).
\bibitem{sundprl} D. Sundkvist, A. Retino, A. Vaivads, S. D. Bale, Phys. Rev. Lett. 99, 025004 (2007).
\end{thebibliography}
\end{document}